# Corona-Enabled Electrostatic Printing for Ultra-Fast R2R Manufacturing of Binder-Free Multifunctional E-Skins

*Long Wang*, *Rui Kou*, *Zhaoru Shang, Chaoyi Zhu, and Ying Zhong*

Dr. L. Wang
Department of Civil and Environmental Engineering
California Polytechnic State University, San Luis Obispo
1 Grand Ave., San Luis Obispo, CA 93407, USA

Dr. R. Kou, Z. Shang
Department of Structural Engineering
University of California, San Diego
9500 Gilman Drive, La Jolla, CA 92093, USA

Dr. C. Zhu
Department of Materials Science and Engineering
Carnegie Mellon University
5000 Forbes Avenue Pittsburgh, PA 15213, USA

Dr. Y. Zhong
Corresponding author: E-mail: yingzhong@usf.edu
Department of Mechanical Engineering
University of South Florida
4202 E. Fowler Avenue, Tampa, FL 33620, USA

* These authors contributed equally to this work

**Abstract**

As essential components in intelligent systems, printed soft electronics (PSEs) are playing crucial roles in public health, national security, and economics. Innovations in printing technologies are required to promote the broad application of high-performance PSEs at a low cost. However, current printing techniques are still facing long-lasting challenges in addressing the conflict between printing speed and performance. To overcome this challenge, we developed a new corona-enabled electrostatic printing (CEP) technique for ultra-fast (milliseconds) roll-to-roll (R2R) manufacturing of binder-free multifunctional e-skins. The printing capability and controllability of CEP were investigated through parametric studies and microstructure observation. The electric field generation, material transfer, and particle amount and size selecting mechanisms were numerically and experimentally studied. CEP printed graphene e-skins were demonstrated to possess outstanding strain sensing performance. The binder-free feature of the CEP-assembled networks enables them to provide pressure sensitivity as low as 2.5 Pa, and capability to detect acoustic signals of hundreds of hertz in frequency. Furthermore, the CEP technique was utilized to pattern different types of functional materials (e.g., graphene and thermochromic polymers) onto different substrates (e.g., tape and textile). Overall, this study demonstrated that CEP can be a novel contactless and ultrafast manufacturing platform compatible with R2R process for fabricating high-performance, scalable, and low-cost soft electronics.

## 1. Introduction

Printed soft electronics (PEs) are playing crucial roles in public health, national security, and economic stability, as they are the essential components in internet of things (IoT) systems providing social distancing,[1] health monitoring,[2,3] real-time diagnosis,[4,5] drug delivery,[6] human-machine interaction,[7] timely treatment, [8] and so forth. The market of PEs is projected to grow from $7.8 billion in 2020 to $20.7 billion by 2025 with a compound average growth rate (CAGR) as high as 21.5%.[9] In particular, printed soft electronics (PSEs) have flourished in recent years and opened a new path for comfortable and personalized ways to realize the above functions.[5,10,11] Innovations in their manufacturing technologies are required to promote the broad application of high-performance PSEs.

However, current printing techniques are facing challenges in addressing the conflicts between printing speed and performance. For instance, contact printing techniques such as screen,[12,13] flexographic, [14,15] gravure [16,17] are relatively easy to realize industrial roll-to-roll (R2R) manufacturing and achieve high printing speed. But they suffer from drawbacks of wearing off of masks, a large amount of waste materials, and low resolution.[12,13,16–18] To address these issues, non-contact photolithography,[19,20] laser direct writing,[21,22] and inkjet printing[23,24] with improved resolution have become popular alternatives. However, their throughputs are much lower than the contact printing techniques because of the poor compatibility with R2R systems or speed limitations in ink feeding. [19~24]

Furthermore, almost all the above techniques require using passive polymer binders for material transfer. However, the usage of binders introduces many limitations, including 1) curing/drying of polymer additives often takes long time and high temperatures, leading to increased manufacturing cost and limited material options;[18] 2) low melting/softening temperature of polymer additives limits the high-temperature applications of PSEs;[25] and 3) insulation and encapsulation of passive materials may compromise the sensitivity of nanomaterial networks.[26] Thus, it is highly desirable to eliminate or reduce the use of passive polymer binders during the printing process.[27] This work is reporting an innovative new printing technique corona-enabled electrostatic printing (CEP) to overcome the above limitations. In CEP, the to-be-printed substrate is place in between the material and corona discharge (CD) in a contactless way. The material transfer in the CEP process is realized by the contactless electrostatic attraction process activated and controlled by the electric field generated by CD. It enables the ultra-fast transfer and patterning of various types of materials with advantages of contactless, binder-free, and ultra-fast for mass production. Firstly, CD is an electrical discharge caused by the ionization of air surrounding a sharp conductor carrying a high voltage.[28–32] It means that the electrode and the to-be-printed substrate is not in contact. Compared with the conventional approaches of utilizing electric field to assemble materials,[33] CEP eliminates the limitations in manufacturing flexibility, material choices, throughputs, and assembly directions through a contactless way; it prevents the abrasion of masks; and it provides more flexibility in coupling electric field with other the material transfer mechanisms. Secondly, the electrostatic force in CEP eliminates the use of passive polymer additives as the media to realize material transfer. It avoids high temperature and long-time drying of binders, as well as the sensitivity reduction caused by passive polymer encapsulation. It also provides more flexibility in choosing the materials, media, and substrates. Thirdly, CEP is an ultra-fast process which finishes material transfer within milliseconds. CD is a well-developed technology that has been widely used to provide surface treatment on polymer films in an R2R manner with a remarkable treatment line speed of up to 500 m/min.[34] The area covered by one single corona discharge wire or needle can be as wide as almost 10 cm. And the length of the wire can reach meter range. Thus, combined with the rolling speed in the R2R process, the printed area by CEP can potentially reach hundreds of meter squares within one minute, making it a great solution for ultra-fast mass production. Plus, the current of CD is as low as tens of microamp, meaning it is a safe and low power consumption process.

To validate the advantages of the CEP technique, in this paper, we systematically investigated its printing capability, controllability, working mechanism, and the performance of the printed sensors. Specifically, CEP was utilized to print different types of materials, including graphene and thermochromic (TC) materials to demonstrate the broad material options of CEP. The influences of corona parameters were studied to understand their controllability over the printed microstructure. Experiments and simulations were conducted to understand the electric field formation and material transfer mechanism. The properties of printed graphene-based strain sensors were tested, including their sensitivity, repeatability, and response to different types of strains, including acoustic vibration. The responsive mechanism of CEP printed binder-free microstructures to applied strain was analyzed. Finally, the CEP R2R manufacturing and the temperature response of TC material-based CEP temperature sensors were demonstrated.

## 2. Results and Discussion

### 2.1 CEP Setup and Printed Microstructures

Figure 1a is the schematic of the basic setup and material transfer behavior of a CEP system with a tungsten needle as the electrode (wire electrode can cover larger area). From top to bottom, the system is formed by the discharge needle (supported by high voltage power supply), the to-be-printed substrate, an optional mask for patterning, and the to-be-printed nanomaterials. Once the corona is turned on, the black color graphene powders in the bottom container were attracted to the substrate instantly. The printed microstructure of samples printed by different corona voltages was observed by SEM. It was discovered that the CEP procedure can be controlled by the corona discharging voltage and is selective to different particle sizes. Specifically, comparing Figure 1b printed by 15 kV corona voltage with Figure 1c printed at 25 kV, and based on the statistics in Figure 1d, an increase of size and number of particles was observed with the increase of corona voltage. The statistic of the particle distribution in Figure 1e shows that with the increase of the voltage, the percentage of smaller particles decreases, while the percentage of larger particles increases. It means that higher voltage tends to attract more larger particles. With the voltage increase, the percentage of the substrate area covered by the printed materials also increases, which is ~61% for 15 kV and ~77% for 30 kV (Figure 1e inset). For all the above tested CD voltages, over 93% of printed particles are less than 50 μm. Thus, by controlling the corona voltage, it is possible to control the selected particle size distribution and amount of the materials. The detailed material transfer mechanism will be elaborated in the next section. Figure 1f is a demonstration showing the substrate printed by CEP is not only limited to flat surfaces. The electrostatic force can also push the particles into the complex 3D structure of non-woven fabrics, expanding the printable substrate choices.

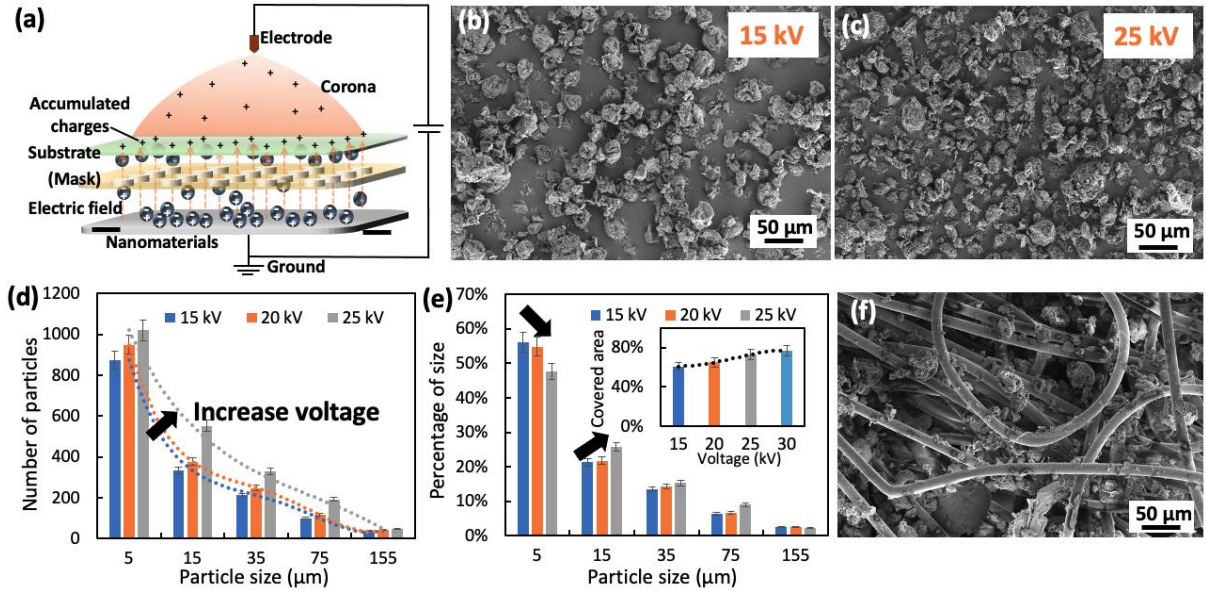

**Figure 1** Corona electrostatic printing (CEP). a) Schematic of the basic needle CEP system. b, c) SEM image of graphene particles attached onto the substrate after CEP of 15 kV and 25 kV. d) Number of particles with different size ranges attached onto the substrate by corona voltages of 15 kV, 20 kV and 25 kV. e) Particle size distribution in percentage by CEP of different voltages; the inset is the percentage of the covered area on the substrate. f) SEM image of graphene printed on non-woven fabric.

## 2.2 CEP Material Transfer Mechanism

As elaborated in Figure S1, the material transfer process is: 1) corona generated, 2) electric field formed between the substrate and the to-be-printed nanoparticles, 3) nanoparticles get polarized by the electric field, 4) nanoparticles attracted to the substrate above it. To realize successful material transfer, the upwards attraction electrostatic force $\vec{F}_e$ needs to overcome the downwards gravity $\vec{G}$. That is

$$\vec{F}_e > \vec{G} = \rho V \vec{g} \tag{1}$$

where, $\rho$ is the density of the particle, $V$ is its volume, and $\vec{g}$ is acceleration of gravity. The attraction electrostatic force is [35–37]

$$\vec{F}_e = Q\vec{E}_{out} \tag{2}$$

where, $\vec{E}_{out}$ is the electric field applied at the outer surface of the particles, $Q$ is the induced charges on the particles. It is obvious that the electric field strength and induced charges are two major factors determining the static force applied on the material. For the same type of material with the same conductivity to obtain similar density of induced charges, the electric field is the master factor. The higher electric field may allow the CEP process to transfer larger particles. Below, we are discussing the formation and controlling mechanism of the electric field in CEP and its dynamic material transfer and particle size selection mechanism.

*2.2.1 Electric Field Formation and Electric Induction Mechanisms*

The electric field distribution of CEP is different compared with conventional CD process as the insert of the to-be-printed substrate accumulates charges and enhances the electric field between the to-be-substrate and the material. Finite element analysis (FEA) with COMSOL Multiphysics was conducted to understand the formation and control mechanism of the electric field during CEP (see more details in Supporting Information). Here we use negative corona as an example for the detailed discussion. Figure 2a displays the dimension of each component in the CEP system. Figure 2b compares the charge density for the corona system with and without the inserted polymer film as the to-be-printed substrate. It can be clearly observed that adding the polymer film can significantly increase the charge density in the area underneath it. It leads to a significant increase of the electric field underneath the polymer film compared in Figure 2c. Without the inserted film, the electric field applied on the bottom area is just around 500 kV/m. However, it increased to

~800 kV/m for -15 kV corona, ~1100 kV/m for -20 kV corona, and ~1400 kV/m for -25 kV (Figure 2d).

The formation mechanism of this enhanced electric field is an interesting physical process. During corona discharging, electrons avalanche occurs around the high voltage sharp needle, colliding with the neutral molecules in the surrounding air and forming a large number of negative ions. The electrical field between the discharge electrode and the grounded electrode pushes the generated negative ions downwards by electrostatic force. Since the highly insulating polymer film locates in between the discharge needle and the ground electrode, the negative charges get accumulated at the top surface of the polymer film until it reaches the maximum value. The accumulated charges form another electric field, which weakens the overall electric field above the film, but enhances the field underneath it. Therefore, the special configuration of the CEP which inserts the polymer film in between the corona and the ground electrode creates a highly enhance electric field, which is beneficial for applying strong electrostatic force on the materials in it and motivating them to move towards the substrate. In addition, the enhanced filed can be formed very quickly. As indicated in Figure 2e, the electric field under the polymer film can reach the equilibrium state within 10 ms, which is beneficial for ultra-fast material transfer.

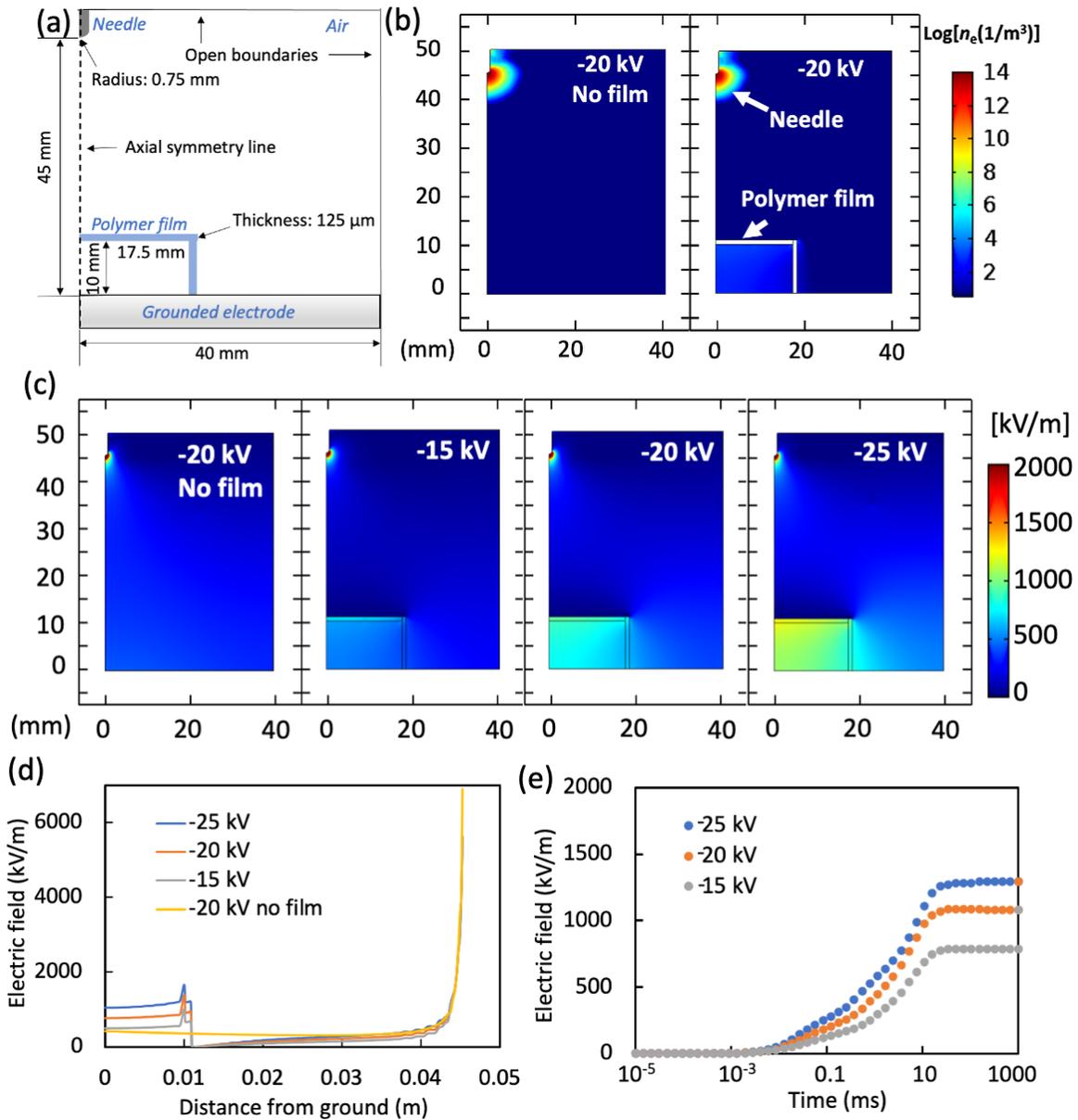

**Figure 2** FEA simulation of the electric field distribution in the CEP setup. a) The configuration of the CEP system used in the FEA simulation with key parameters the same with the experiments. b) Charge density distribution in the corona discharge system with and without the polymer film when corona voltage was set as -20 kV. c) Electric field distribution in the system at corona voltages of -15 kV, -20 kV (with and without film), and -25 kV. d) The electric field distribution

along the central line in figure a). e) Electric field strength under the polymer film increases with time.

*2.2.2 Dynamic Material Transfer and Particle Size Selection Mechanisms*

The discussion above reveals the formation mechanism of the electric field without considering the dynamic attachment of the to-be-printed materials on the substrate during the printing process. In fact, with more and more materials covering the polymer film, the electric field underneath the polymer film dynamically changes. It is because the attracted materials carry charges to the film, it neutralizes the accumulated charges stored on the film. This dynamic process affects the material transfer process and result. Figure 3a is the simulated electric field distribution with 20%, 50% and 100% of the substrate covered by conductive material. It is observed that with more area covered, the strength of the electric field reduces. Figure 3b shows the evolution of the average charge density on the polymer films during the first one second of the CEP process. Higher corona voltage leads to higher initial charge density, meaning higher electric field applied on the material. However, all of them quickly reduces to as low as around $0.5 \times 10^{-5}$ C/m$^2$ within 200 ms, and drops down to ~$0.25 \times 10^{-5}$ C/m$^2$ after another 200 ms. As a result, it leads to a rapid reduction of the electric field between the substrate and the material with more material being attached on the substrate, until it is too weak to attract countable amount of materials. It also indicates that the visible material transfer happens very fast within hundreds of milliseconds.

The attracted particle amount and selected particle size also varies with the initial corona voltage and the dynamic evolution of the electric field. Movie 1 and Figure 3 c~h and are the movie and snapshots of the simulated dynamic material transfer process, which explains the dynamic material transfer mechanism in detail. In the same electric field, smaller particles with lower mass/gravity can obtain larger velocity to reach the substrate faster. During the first 5 ms (Figure 3c), particles smaller than 5 μm first reached the upper substrate firstly. At the 10$^{th}$ ms (Figure 3d), particles in the range of 5 μm to 10 μm start to arrive at the substrate. At the 50$^{th}$ ms (Figure 3e), the arriving particles size increased to over 20 μm. At the 100$^{th}$ ms (Figure 3f), the size of the arriving and departing particles are similar, including all size ranges from less than 5 μm to over 100 μm. But at the 200$^{th}$ ms (Figure 3g), the size of the moving particles reduced to around half of that at the 100$^{th}$ ms. This is because of the ongoing reduction of the electric field strength. On the 500$^{th}$ ms (Figure 3h), the moving particles are reduced back to less than 10 μm. Thus, the material transfer process is dynamic with more materials getting attached onto the substrate and weakens the field. Controlling the initial corona voltage and the dynamic process can provides us the capability of selecting the particle amount and size range.

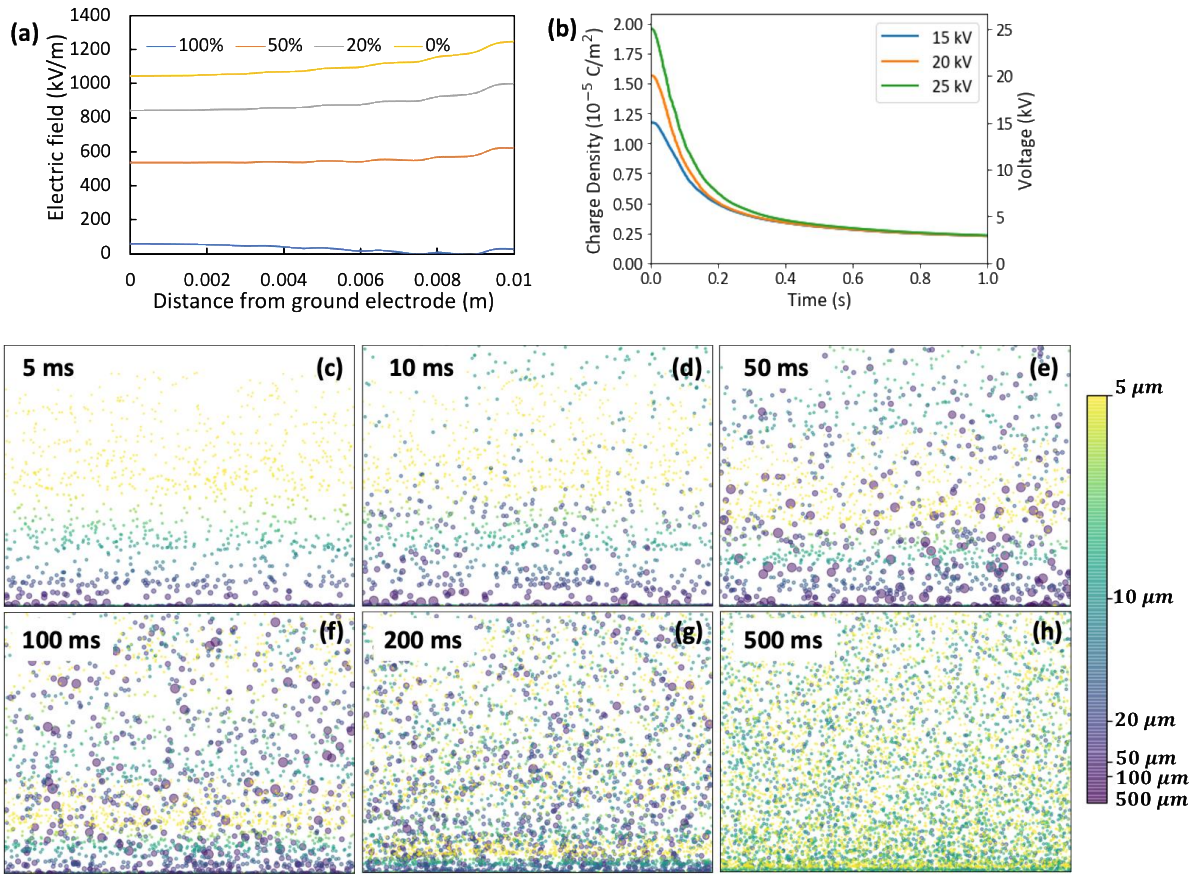

**Figure 3** Simulated dynamic electric field evolution and material transfer process. a) Electric field evolution with 0, 20%, 50%, and 100% of the area covered with the conductive material. b) Charge density between the substrate and the ground electrode evolves with time in the first second of CEP at different voltages of 15 kV, 20 kV, and 25 kV. c~h) Snapshots of the dynamic material transfer process at different times of 5 ms, 10 ms, 50 ms, 100 ms, 200 ms, and 500 ms for 15 kV CEP; the color of the dots represents their sizes, the upper side in each image represents the substrate, and the bottom side represents the material resource side.

Figure 4 gives the displacement, velocity, and acceleration of typical individual particles (10 μm, 20 μm, 50 μm, and 115 μm) during the CEP process to explain the material transfer mechanism in more details. Figures 4a~c represent the acceleration, velocity and displacement of four particles with different sizes who start their movement at the 100$^{th}$ ms. Smaller particles can obtain larger initial acceleration as their gravity are smaller (Figure 4a). Therefore, the 10 μm particle can arrive at the substrate within 12 ms; but it took the 20 μm particle 16 ms, and the 50 μm particle 30 ms instead. But no matter what the particle size is, the acceleration drops down with more materials covering the substrate during the CEP process, weakening the electric field and upward attraction force. One extreme phenomenon is that the acceleration and velocity of the 115 μm particle became negative after departing ~90 ms (Figure 4a and b). After traveling upwards for ~0.3 cm, the particle started to dropdown (Figure 4c). As indicated in Figures 4d, starting from the 200$^{th}$ ms, the acceleration of all particles are much lower compared with the ones started from the 100$^{th}$ ms. The initial acceleration of the 10 μm particle was reduced to ~50 m/s$^2$, and 20 m/s$^2$ for the 20 μm particle. The largest activated particle in this time frame is reduced to 50 μm. It requires ~150 ms to arrive at the substrate, with its acceleration and velocity reduced to almost 0 upon arrival (Figures 4e and 4f). For the 500$^{th}$ ms, only particles smaller than ~20 μm can move upward because of the weakening of the field. It takes as long as 130 ms for a 20 μm particle to arrive at the substrate with much lower acceleration and velocity than before (Figures 4g~i).

From the above analysis, it can be concluded that the material attraction process during CEP is a dynamic and selective process. At the beginning, the electric field is strong, and materials begin to move to the targeted substrate at high acceleration and velocity. Smaller particles arrive at the substrate first and larger particles can be attracted. With more material covering the substrate, the electric field weakens, and the size and velocity of the particle reduces till it is negligible after hundreds of milliseconds. The voltage of corona is the major factor controlling the maximum electric force applied on the materials and determining the amount of the attracted particles and the selected particle size range.

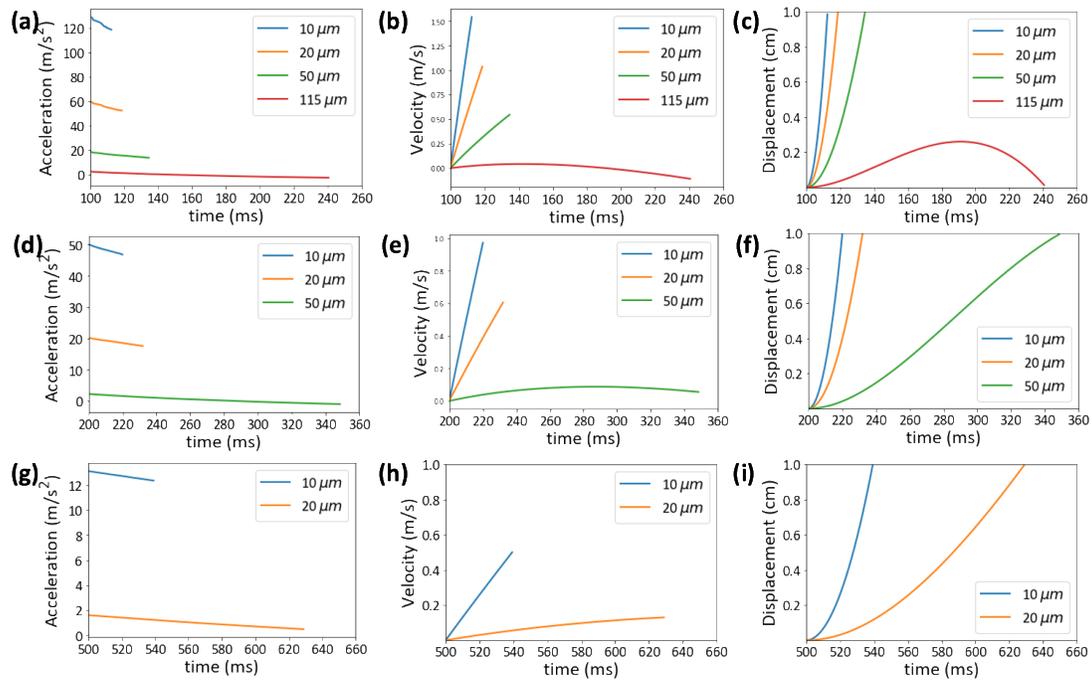

**Figure 4** The acceleration, velocity, and displacement of simulated graphene particles with diameters of 10 μm, 20 μm, 50 μm, and 115 μm at the fabrication time instants of a~c) 100 ms, d~f) 200 ms, and g~i) 500 ms.

**2.3 Sensing Performance of the CEP sensors**

*2.3.1 Strain Sensing*

*1) Electromechanical properties*

The electromechanical performance of the binder-free CEP graphene networks was characterized by conducting cyclic uniaxial tensile tests. The electrical resistance of the graphene-based pattern was simultaneously measured during the loading process. The schematics of the experimental setups for the tensile tests are demonstrated in Figures 5a. Figure 5b shows the representative time history of the normalized change in resistance ($\Delta R_n=(R-R_0)/R_0$) of the CEP sensors (1 cm-wide and 3 cm-long) when they were subjected to the 200-cycle tensile strain patterns (*i.e.*, max strain: 10%; load rate: 1% s$^{-1}$). It can be observed that the resistance of the CEP sensor changed in tandem with the applied strain pattern, which indicates that the CEP-assembled graphene networks were piezoresistive. The inset of Figure 5b demonstrates that the strain sensing response was highly reversible and stable after an initial decay. Here, decay was observed in the beginning 80 cycles, mainly because of the microstructure reconfiguration of the virgin graphene networks when subjected to deformations. After the ripening/training process, the decay was significantly mitigated. Figure 5c plots the normalized resistance change as a function of applied strains overlapped with a least square fitting line, which indicates that the strain sensing response was approximately linear.

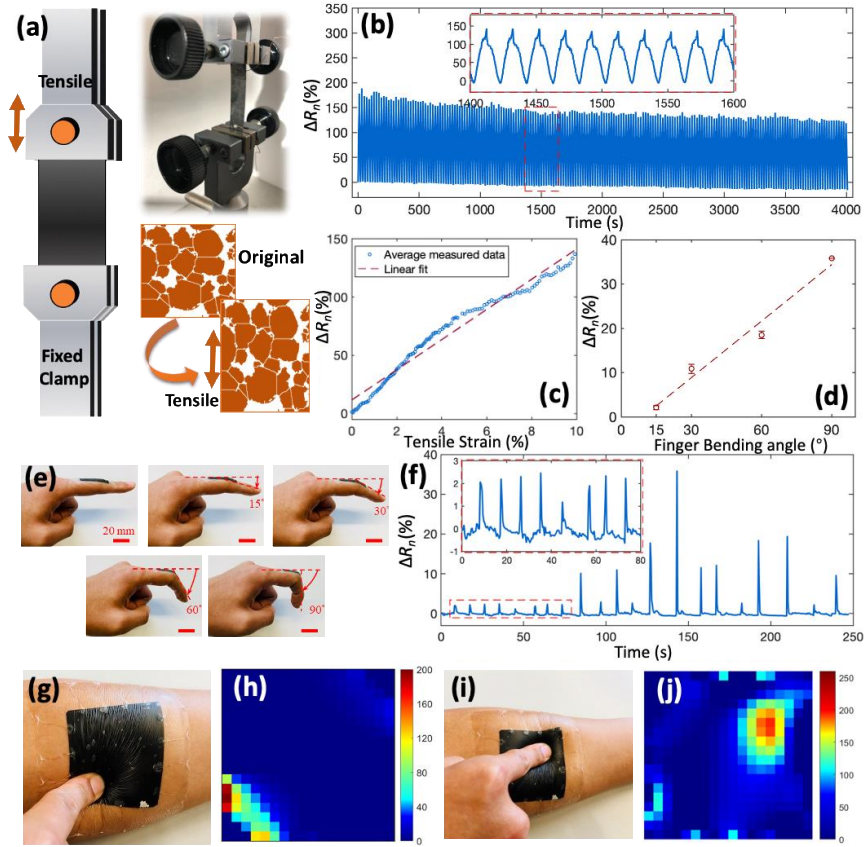

**Figure 5** Electromechanical performance of CEP sensors. a) Schematic and images of the uniaxial tensile test set up. b) Results of the cycling tensile tests. c) Relationship between the normalized resistance change and tensile strain. d~f) Relationship between the normalized resistance change and finger bending angle of 0, 15, 30, 60 and 90 degrees. g~j) Mapping of the location of applied force on human arm.

*2) Human motion monitoring*

The CEP strain sensors were first demonstrated to monitor human motions. In Figures 5d~f, the skin-like CEP sensor was attached on the index finger along its longitudinal direction for monitoring the finger bending degrees (*i.e.*, 15˚, 30˚, 60˚, and 90˚). Since CEP sensors were highly flexible, it could compliantly deform with the finger during bending. The response was highly reversible and repeatable, and the sensing performance exhibited an approximate linear relationship with the finger bending angles, which indicated that the CEP sensors could be used to noninvasively monitor the finger motions.

*3) Spatial strain distribution mapping*

By taking advantage of the scalability of the CEP technique, the graphene-based CEP e-skins could be readily scaled up. In Figures 5g~ j, a 640 × 640 mm$^2$ CEP printed e-skin was attached on the forearm of a human subject. To achieve spatial pressure mapping capability, the CEP e-skin with 4 × 4 boundary electrodes was coupled with the electrical impedance tomography (EIT) measurement scheme and algorithm (see more details in Supporting Information), which could allow one to determine the electrical conductivity/resistivity distribution of the e-skins. At undeformed state, the resistivity distribution of the sensing skin was mostly uniform, indicating an approximately homogeneous electrical property achieved by CEP printing technique. On the other hand, 'hot spots' were identified on the reconstructed resistivity maps upon pressure was applied on the sensing domain. According to the color bars, hotter color refers to larger increase in the electrical resistivity. Since the CEP e-skin was locally deformed, each reconstructed resistivity map shows a distinct hot spot (*i.e.*, localized increase in resistivity) at the vicinity where the specimen was pressed. It was found that the coupled sensing system could locate the pressure points with relatively high accuracy across the entire area. Therefore, the results indicate the scalable continuous sensing skins could detect and spatially locate the applied contact pressure, which paved the way for their potential applications as large-scale and low-cost artificial smart skins and human-machine interfaces.

*2.3.2 Strain sensing mechanism*

Characterizing the mechanisms of the strain responsive performance of the nanomaterial-based strain sensors has attracted increasingly more attention. Different computational models and numerical simulation strategies, such as tunneling effect- and Monte Carlo-based methods, among others, have been developed to analyze the strain sensing mechanisms of the strain sensors.[38–41] However, few reported efforts have associated the piezoresistive behaviors with the actual microstructures of the sensing materials. This is mainly due to the challenge in imaging the loading effects on the microstructures and the encapsulation of passive binders.

For the binder-free CEP-fabricated microstructures, as it eliminates the uncontrollable encapsulation of the passive binder, it was hypothesized that the observed piezoresistive response was mainly resulted from the deformation-induced alternation in the microstructures of the binder-free networks. In particular, the applied tensile strains could decrease the global electrical conductivity of the graphene patterns through disturbing the graphene network connections. However, the disturbed network connections could restore to their initial state when the strains were unloaded, which was demonstrated by the aforementioned reversible electromechanical response. In order to validate the tension-induced change in the microstructures of the graphene networks, the digital image correlation (DIC) technique has been employed to map the full-filed displacements and strains of the graphene networks in a non-contact manner. The DIC technique, as an optical metrology, has been used to quantify the deformations of objects based on digital image processing and numerical computing.[42,43] Typically, the surfaces of objects need to incorporate laser or white-light speckle patterns that could transfer the displacement information to the DIC method. Here, the DIC analysis was conducted based on the *in-situ* microscopic optical images of the graphene networks, where the graphene particles were directly used as the speckles for tracing the displacements of the networks. The *in-situ* optical images in Figure 6 were taken from the graphene patterns fabricated using a charging voltage of 25 kV. Figures 6a~d show the horizontal displacement fields of the graphene network when it was subjected to 5%, 10%, 15%, and 20% uniaxial tensile strains, respectively. The moving clamp (set up on the right-hand side of Figures 6a~d) applied uniform displacement to the graphene network. In addition, strains were calculated by displacements (evaluated using DIC) divided by the initial dimensions of the images, and the horizontal strain ($\varepsilon_x$) maps are shown in Figures 6e~h, corresponding to Figures 6a~d, respectively. Since the graphene particles were assembled with no binder, it essentially isolated some of the graphene particles from each other. The DIC-based evaluated tensile strains relatively accurately corresponded to the practically loaded deformations.

In this study, finite element analysis (FEA) was performed based on actual microstructures of the graphene networks captured using the aforementioned *in-situ* microscopic optical imaging method, which was demonstrated capable of characterizing the strain-caused microstructural changes in the graphene networks (see more details in Supporting Information). Figures 6i~l show the electric potential field distributions across the graphene networks when subjected to 5%, 10%, 15%, and 20% tensile strains, respectively. It was found that as the graphene particles became more isolated, higher electric potential would be generated, which indicated an increase in bulk electrical resistance of the graphene network. Thus, the microstructure-based FEA further validated that the deformation-induced reconfiguration of graphene network could lead to the experimentally observed strain sensitive performance.

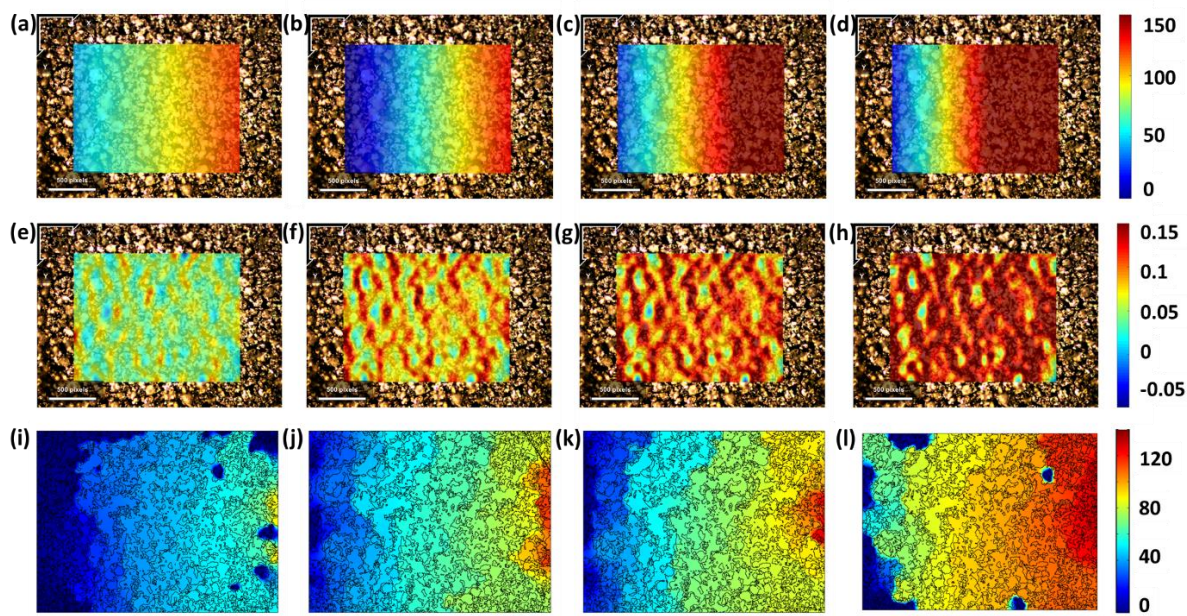

**Figure 6** Graphene network's microstructural change under strains and numerically simulated electromechanical response. a~d) Displacement fields overlapped with the optical microscopic images of the graphene network when subjected to 5%, 10%, 15%, and 20% uniaxial tensile strains, respectively. e~f) DIC-measured horizontal strain distributions corresponding to a~d, respectively. i~l) Numerically simulated electrical potential distributions when the graphene network was under 5%, 10%, 15%, and 20% uniaxial tensile strains, respectively.

## 2.4 Multifunctional CEP sensors

### 2.4.1 Acoustic signal detection

The binder-free nature of CEP allows the nanomaterial network to deform without the restriction from the encapsulating binders, leading to enhanced sensitivity (down to 2.5 Pa as elaborated in Figure S2). It is possible for the materials to vibrate actively to accurately detect the frequency and intensity of the acoustic signal. Figure 7a demonstrated the capability of CEP graphene sensor prototypes to detect a range of musical notes from Do to Ti. Assisted with fast Fourier transform

(FFT) analysis, Figures 7b and 7c are indicating the capability of CEP graphene sensors detecting the frequencies of acoustic signals, such as the demonstrated 50, 100, 150, 200, 250, 300, and 350 Hz (maximum frequency limitation of the multimeter). The capability to detect acoustic signal with higher frequency is possible if the data collecting capability of the multimeter allows. The binder or encapsulation free CEP printed microstructure is responsible for the outstanding vibration capability to detect the acoustic signal. Compared with other flexible acoustic sensors designed for artificial speakers and eardrums, the CEP sensors are manufactured in an ultra-fast way, instead of going through the complicated nano-/micro-patterning procedures, such as photolithography and/or imprinting. Plus, CEP is capable to print those sensors at large scale with a R2R method (Figure S4) and low cost, meaning it can potentially be an affordable manufacturing technique for flexible artificial speakers and eardrums for their broad application.

*2.4.2 Skin-like CEP thermochromic (TC) sensors*

In addition to the strain sensors fabricated with conductive materials, CEP can also print or pattern non-conductive materials and realize more functions as multifunctional flexible sensors. To validate this hypothesis, CEP was demonstrated to print TC particles for temperature sensing application. A mask was inserted between the material and the substrate during CEP. As shown in Figures 7d~g and Movie 2 (more details in Figure S5), CEP-patterned temperature TC patterns changed their colors within seconds when subjected to temperature change. The color change can be monitored by a camera. And by analyzing the intensity and RGB data of the recorded video, the color change of CEP sensors can be digitalized to allow automatic temperature monitoring as demonstrated in Figures 7f and 7g. The CEP technique can potentially mass produce temperature sensors to perform as tattoo-like, wireless, and powerless temperature detectors, which may be widely used in human health monitoring, potentially to track fever symptoms to combat pandemics such as COVID-19.

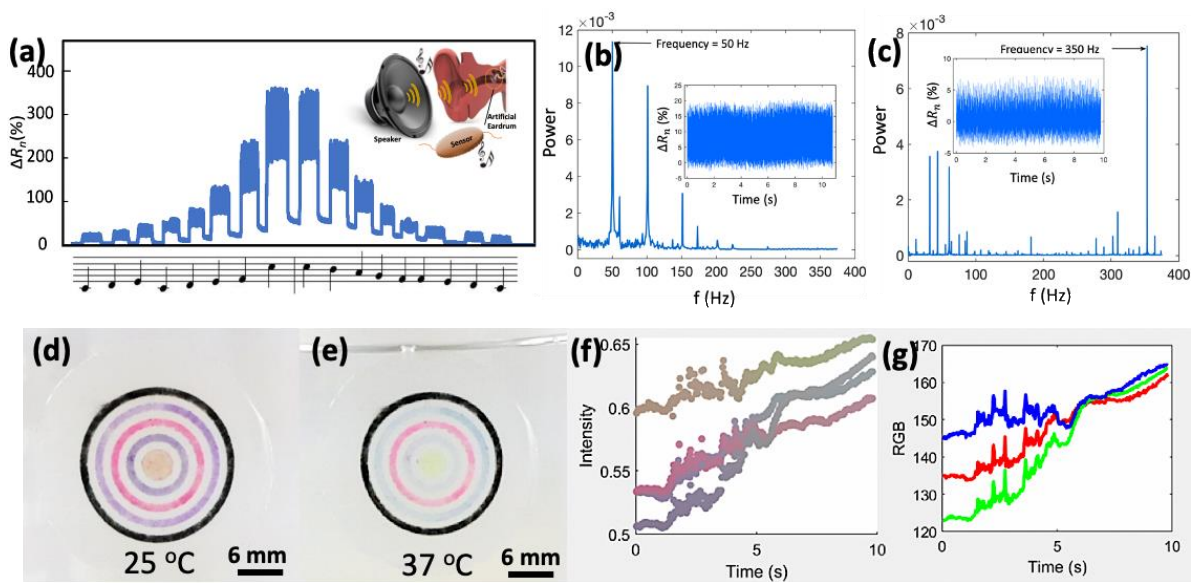

**Figure 7** Demonstrations of CEP printed acoustic and temperature sensors. a) CEP sensor detecting a range of musical notes from Do to Ti. b,c) the result of the CEP printed binder-free graphene sensors detected acoustic signal frequency after FFT analysis. d) The CEP printed pattern with the inside 4 circles printed by thermochromic (TC) materials, and the outermost circle printed by graphene, room temperature. e) The TC patterns changed their color after 37 °C water was pulled into the container which the CEP sensor was attached on. f) The extracted color intensity data of all the four TC circles from the color change video. g) RGB data of the innermost circle.

## 3. Conclusion

In this paper, we reported a new corona-enabled electrostatic printing technique (CEP) for ultra-fast R2R manufacturing of binder-free multifunctional e-skins. Corona discharge (CD) provides a non-contact way to generate the controllable electric field to realize ultra-fast material transfer without utilizing binders as the media. By controlling the corona voltage, the amount of particles and particle size distribution can be manipulated accordingly. Compared with conventional corona discharge, inserting the substrate between the corona and the material enhances the electric field applied on the materials because of charge accumulation on the substrate. The material transfer is a dynamic process which can be accomplished within less than 200 ms. The dynamic electric field evolution during the CEP process enables us to control and select the printed material. The CEP printed sensors were confirmed to possess outstanding strain sensing capability, with the electromechanical relationship linear, reversible and competitive sensitivity of 2.5 Pa. The binder-free microstructure provides an *in-situ* way to understand the strain sensing mechanism caused by the deformation between the network interconnections. The binder-free nature of CEP sensors allows it to possess the capability to detect acoustic signals as it can vibrate sensitively. In addition of graphene, CEP can be utilized to print thermochromic materials for temperature monitoring, and more broader applications by integrating more types of materials together as multifunctional sensors. To summarize, CEP is an ultra-fast printing technique which can be utilized to manufacture highly sensitive binder-free flexible sensors at large scale and low cost through R2R process. The material and substrate choice of CEP can be very broad including conductive and non-conductive materials, and complex 3D structured substrates. As a new advanced manufacturing strategy, CEP will potentially transform the cost structure for large-area and high-performance electronics and enable versatile applications of flexible functional systems.

## 4. Experimental Section

### 4.1 Materials

*4.1.1 Synthesis of Graphene Dry Particles*

Graphene nanosheets were synthesized from graphite powders (-325 mesh, 99.995% pure) using an optimized water-assisted liquid-phase exfoliation (WALPE) technique.[44,45] Here, N-methylpyrrolidinone (NMP, 99% extra pure) and deionized water were mixed at 8:2 ratio for preparing the co-solvent. 5 mg mL$^{-1}$ of graphite powders were mixed with the co-solvent and the mixture was subjected to the bath sonication for 6 h at a temperature between 27 ˚C and 37 ˚C. Then, the dispersion was centrifuged at 3000 rpm for 30 min and the 75% of the resulting supernatant was used to collect graphene. The samples were dried at 80 ˚C for 24 hours to obtain the dry powders.

*4.1.2 Other Materials*

The thermochromic (TC) particles were purchased from Solar Color Dust with three different colors for three temperature ranges of below 27.8 ºC, 27.8 ºC to 35 ºC, and higher than 35 ºC. CNT and TC dry particles was used as received. The substrate for printing was the transparent, breathable and ultra-thin medical tapes obtained from Nitto (XTRATA® Perme-Roll AIR$^{TM}$). The thickness of the tape was ~10 μm, and its mechanical property was compatible to human skin. The fibrous textile is the polypropylene nonwoven clothes from TechniCloth.

**4.2 CEP Printing Process**

For the lab-scale CEP setup, as schematized in Figure 1a, the to-be-printed substrates were cut into needed sizes for the demonstration of sensors with different functions. Graphene or TC dry particles were uniformly placed in polystyrene petri dishes (60 mm in diameter and 15 mm in height). The tape was fixed on the lid of the container with the adhesive side exposed to the powder. Then this set was placed 3 cm underneath the discharge tungsten needle (tip ~0.1 mm) of the corona charging system, consisting of the high voltage power supplier from Glassman Co. Lt. which can provide adjustable voltages up to $\pm 40$ kV. Even though, the material transfers visually happened and finished immediately once the corona was turned on. The CD was kept on for 5 sec for the purpose of consistency and controllability. After the CEP process, the samples were uniformly and gently blasted with air for 5 seconds. Silver paste was utilized to connect the stainless thin conductive yarn (from Adafruit Inc.) as electrodes onto the sensors to collect signal. The second layer of the skin-like Nitto tape was adhered onto the printed samples to form the flexible ultrathin sensor. For patterning, laser-cut patterned masks were attached on the adhesive side of the tapes, and then removed afterwards the printing process.

### 4.3 Microstructure Characterization and *In-situ* Monitoring

The microstructure of the printed powders was characterized by optical microscopy and scanning electron microscope (SEM). The particle distribution was analyzed with Image J software. To monitor the microstructure evolution of the CEP network under various strains, the printed samples were hold with a home-made stretching device as shown in Fig. S??. Strains of 5%, 10%, and 20% were applied to the samples under *in-situ* optical microscopy observation. The real-time microstructure evolution was compared with the electrical property change during the increase of the strain to analyze the working mechanism of CEP strain sensors. The electrical signals were collected by Keysight multimeter 334465A.

### 4.4 CEP Electric Field Simulation

COMSOL Multiphysics 5.3a (Application Gallery ID 14035: Atmospheric Pressure Corona Discharge) was employed to simulate the CEP process using finite element analysis (FEA). Figure 2a shows a two dimensional (2D) axial symmetrical cross section of the FEA model. In this setup, the needle has a length 4 mm and a tip radius of 0.75 mm. The distance between the needle tip and the grounded electrode is 45 mm. The 10-μm thick skin-like senor substrate is attached on a 125 μm polystyrene film with a diameter of 35 mm, and they are fixed 10 mm above the grounded electrode. More details about the simulation are given in the Supporting Information.

### 4.5 Dynamic Material Transfer Simulation

The numerical simulation of the dynamic material transfer process was carried out with Python 3.7. The system configuration and particle size distribution are in accordance what mentioned in above FEA and experiments. In order to simplify the calculation, the particles were in cubic shape. The polymer films were preset with different surface charge amount $Q_0$, and the initial electric voltage varied as 15 kV, 20 kV, and 25kV. During the simulation, the acceleration, velocity and displacement from the grounded electrode of the particles were recorded every 0.2 ms. For simplicity, particles reaching the polymer substrate or material resource plane were recorded and deleted from the system after their charges were transferred to the substrate.

### 4.6 CEP Sensor Properties Characterization

*4.6.1 Spatial pressure sensing tests*

To perform the EIT measurement on the CEP sensors, a customized data acquisition system was employed to automatically interrogate and record the boundary voltages of the specimens. To be specific, a Keithley 6221 current generator was used to inject 1 mA of DC across a preselected pair of boundary electrodes. The current generator was interfaced with a Keysight 34980A multifunctional switch that was controlled using a customized MATLAB program for automatically switch the pairs of electrodes. In addition, a built-in DMM of the switch could measure and record voltages between the other pairs of electrodes. For each undeformed and deformed case, the whole set of voltage measurements were used as inputs to the EIT algorithm for reconstructing the resistivity distribution of the graphene-based sensing network.

*4.6.2 Characterization of multifunctional CEP sensors*

*1) Acoustic signal detection*

CEP sensors were attached onto a speaker, with a 2 cm gap between the sensor and the bottom of the speaker. Acoustic signals with different frequencies were generated by a phone APP *f* Generator.

*2) Temperature monitoring*

The TC materials were firstly directly treated with corona with opposite polarity with the CEP corona. The voltage was set as 10 kV and the corona treatment time was 3 sec. After which, the TC materials went through the CEP. A laser-cut mask was attached on the adhesive side of the substrate to form patterns. After CEP, sensors were tightly attached on transparent glass beakers. Water of 37 °C was pulled into the beakers with the camera on to monitor the color change. The RGB signal was analyzed by MATLAB to obtain digital data of the color change. Video footage data (.mp4) was converted into an object in Matlab using the VideoReader function. In this study, the VideoReader object contained an RGB24 format video shot at 30 fps. To conveniently analyze time-resolved information corresponding to each ring of interest, uniformly sampled pixels, 7.35° between two pixels on the same ring and 50 pixels per ring, were determined for every ring in every frame to obtain average intensity and RGB color profiles. Besides, the intensity profiles, a weighted average of the RGB channels with a range from zero to one (rgb2gray function), were colored by the hue of the ring and plotted as a function of time. The average RGB channels were also plotted as a function of time but colored by their corresponding colors.

**Supporting Information**

Supporting Information is available from the Wiley Online Library or from the author.

**Acknowledgements**

The authors acknowledge the research grant support of the Startup Funding for Dr. Ying Zhong from University of South Florida. The authors also acknowledge the support from Dr. Yu Qiao and Dr. Kenneth Loh at the University of California, San Diego for the corona system, material supplies and the EIT data acquisition hardware systems.

**Conflict of Interest**

The authors declare no conflict of interest.